\documentclass[a4paper,fleqn,usenatbib,useAMS,usedcolumn]{mnras}
\usepackage[T1]{fontenc}
\usepackage{ae,aecompl}
\usepackage{amssymb,amstext,amsfonts}
\usepackage[fleqn]{amsmath}
\usepackage{graphicx}
\usepackage{subfig}
\usepackage{dcolumn}
\usepackage{multirow}
\usepackage{hyperref}

\usepackage{etoolbox}
\makeatletter
\patchcmd\@combinedblfloats{\box\@outputbox}{\unvbox\@outputbox}{}{%
  \errmessage{\noexpand\@combinedblfloats could not be patched}%
}%
\makeatother

\usepackage{tabularx}

\newcommand{\de}{\mathrm{d}}
\newcommand{\be}{\begin{equation}}
\newcommand{\ee}{\end{equation}}
\newcommand{\bi}{\begin{itemize}}
\newcommand{\ei}{\end{itemize}}

\newcommand{\bml}{\begin{mathletters}}
\newcommand{\eml}{\end{mathletters}}
\newcommand{\bl}{{\vec \lambda}}

\newcommand{\bd}{{\vec d}}
\newcommand{\binfo}{{\cal I}}

\newcommand{\model}{{\cal H}}
\newcommand{\GOB}{\widetilde{G}_{\mathrm{OB}}}
\def\ltsima{$\; \buildrel < \over \sim \;$}
\def\simlt{\lower.5ex\hbox{\ltsima}}
\def\gtsima{$\; \buildrel > \over \sim \;$}
\def\simgt{\lower.5ex\hbox{\gtsima}}
\title[Tests of GR with binary pulsars]{
On tests of general relativity with binary radio pulsars}
\author[W.\ Del~Pozzo \& A.\ Vecchio]{\parbox{\textwidth}{W.~Del~Pozzo$^{1}$\thanks{E-mail: wdp@star.sr.bham.ac.uk} \& A.~Vecchio$^{1}$}\vspace{0.2cm}\\
\parbox{\textwidth}{$^{1}$School of Physics \& Astronomy, University of Birmingham, Edgbaston, Birmingham, B15 2TT, UK}\\
}

\begin{document}

\date{\today}

\pagerange{\pageref{firstpage}--\pageref{lastpage}} \pubyear{2016}

\maketitle

\label{firstpage}

\begin{abstract}
The timing of radio pulsars in binary systems provides 
a superb testing ground of general relativity.
Here we propose a Bayesian approach to carry out these tests, and a relevant
efficient numerical implementation, that has several conceptual and practical advantages with respect to traditional methods based on least-square-fits that have been used so far:
(i) it accounts for the actual structure of the likelihood function -- and it is not predicated on the Laplace approximation which is implicitly built in least-square fits that can potentially bias the inference -- (ii) it provides the ratio of the evidences of any two models under consideration as the statistical quantity to compare different theories, and (iii) it allows us to put \emph{joint} constraints from the monitoring of multiple systems, that can be expressed in terms of ratio of evidences or probability intervals of global (thus not system-dependent) parameters of the theory, if any exists. Our proposed approach optimally exploits the progress in timing of radio pulsars and the increase in the number of observed systems.
We demonstrate the power of this framework using simulated data sets that are representative of current observations.
\end{abstract}


\section{Introduction}
\label{sec:Introduction}

Radio pulsars observations allow some of the most exquisite tests of two-body dynamics in general relativity (GR) (\cite{Eardley:1975,DamourEspositoFarese:1992,DamourEspositoFarese:1996, KramerEtAl:2006, Foster:2007}; see also \cite{Stairs:2003, Will:2014} for recent reviews). They rely on the monitoring of the time of arrival of the radio pulses of these rotating neutron stars orbiting a companion compact object, a neutron star or a white dwarf \citep{BellEtAl:1996,BhatEtAl:2008,ShaoWex:2012,FreireEtAl:2012,Shao:2014,YagiEtAl:2014}. With the ever-increasing sensitivity, observational baseline and sky coverage of radio observations and with the advent of new major telescopes, such as the Five hundred meter Aperture Spherical Telescope (FAST) and the Square-Kilometre-Array (SKA), one can expect the discovery of new (possibly extreme) systems and the further increase of the accuracy in monitoring existing ones. As a consequence, progressively more stringent constraints will be placed on the two-body dynamics.

The time of arrival (TOA) at the Earth of the radio pulses emitted by pulsars in binary systems depends on the Keplerian parameters of the binary as well as on a set of post-Keplerian parameters. The relations of the latter to the binary parameters depend on the specifics of the theory of gravity. Traditionally, the parameters values are obtained by means of a least-square fit. Tests of the gravity are then carried out by checking consistency relations of over-determined parameters of the theory, and/or recovering probability intervals of a given parameter which is consistent with the theory ({\it e.g.} \cite{BurgayEtAL:2003, KramerEtAl:2006}; see however \cite{ZhuEtAl:2015} for an hybrid Monte Carlo approach).
This approach has proven very successful so far. Nonetheless, it suffers from conceptual and practical limitations that hinder the exploitation of the full information content of these exquisite observations, present and future. Specifically, the fit implicitly assumes a quadratic approximation of the likelihood function, which in principle does not hold (see \cite{Vigeland:2014} for further discussions).
Moreover, all the tests hitherto performed are done on a system-by-system basis: they do not take advantage of the fact that much more stringent constraints can be put by combining information across multiple systems.

In this letter we introduce an approach that addresses these shortcomings working in the framework of Bayesian inference: a test to compare different theories of gravity {\it and} to combine information from observations of several systems. We also provide a numerical implementation based on nested sampling that is both accurate and efficient. We illustrate the power of our approach by using a set of simulated data modelled around existing observations.

\section{Method and models}
\label{sec:ModelMethod}

The radio pulses' TOAs at the Solar System Barycenter, $\tau$, are related to the emission time, $t$, at the location of the pulsar by $t = \tau + \Delta(\vec{\lambda})$, where we define $\Delta(\vec{\lambda})$ as the total contribution of the delays that depend on a set of unknown parameters $\vec{\lambda} = \{\vec{\theta}, \vec{\lambda}_\mathrm{K}, \vec{\lambda}_\mathrm{pK}\}$. For convenience, we divide the parameters in two categories: non-orbital parameters $\vec{\theta}$ -- position, proper motion, propagation through the interstellar medium, white and red noise parameters, etc. -- and orbital parameters $\{\vec{\lambda}_\mathrm{K}, \vec{\lambda}_\mathrm{pK}\} $. In any relativistic theory of gravity, the latter can be further divided into 5 Keplerian parameters, $\vec{\lambda}_\mathrm{K}$, which 
%, that 
are connected to the Keplerian description of the orbital dynamics, and a set of post-Keplerian parameters $\vec{\lambda}_\mathrm{pK}$ that encode the higher order corrections to the Keplerian dynamics, and are specific to the theory of gravity under consideration.
%due to the theory of gravity. 
Observations of pulsars according to any relativistic theory of gravity are described by the same set of parameters $\vec{\theta}$ and $\vec{\lambda}_\mathrm{K}$ but different post-Keplerian parameters $\vec{\lambda}_\mathrm{pK}$.

In order to illustrate our proposed approach, in this Letter we consider three specific models for the orbital dynamics that are commonly considered in the analysis of pulsar timing data, based on the seminal work by \cite{DamourDeruelle:1986} and~\cite{DamourTaylor:1992}. Currently, only 5 post-Keplerian parameters are needed to describe the TOAs for the best systems -- for the double pulsar, measurements of the relativistic spin-orbit precession rate come from radio eclipse models~\citep{BretonEtAl:2008} -- but our method is general and can include the appropriate additional parameters required to describe future data. We consider the Keplerian parameter vector $\vec{\lambda}_\mathrm{K} = \{P_b, x_p, e, \omega_0, T_0\}$, where $P_b$ is the orbital period, $x_p$ the semi-major axis along the line of sight, $e$ the orbital eccentricity, $\omega_0$ the longitude of the periastron at a reference time $T_0$, the epoch of the periastron. The post-Keplerian parameters $\vec{\lambda}_\mathrm{pK}$ are:
$\dot{\omega}$, the average rate of the periastron advance, $\gamma_p$, the amplitude of delays in arrival of pulses caused by the varying effects of the gravitational redshift and time delation as the pulsar moves in its elliptical orbit at varying distances from its companion, $\dot{P}_b$, the rate of change of the orbital period, and $r_p$ and $s_p$, the ``range" and ``shape", respectively, of the Shapiro time delay of the pulsar signal as it propagates through the curved space-time region near the companion. In any {\it specific} theory of gravity, $\vec{\lambda}_\mathrm{pK}$ are (theory-dependent) functions of $\vec{\lambda}_\mathrm{K}$ and the unknown binary component inertial masses -- $m_p$, the mass of the pulsar and $m_c$ the mass of the companion -- the equation of state of the neutron stars (and in some cases the two angles that describe the pulsar spin axis).

In GR, the five post-Keplerian parameters are related to $\vec{\lambda}_\mathrm{K}$ and the two masses by the following equations~\citep{KramerWex:2009}:
\begin{align}\label{eq:gr-ppk}
\dot{\omega}&=
3T_\odot^{2/3}n_b^{5/3}\frac{1}{1-e^2}(m_p+m_c)^{2/3}\,,
\\
\gamma_p &=
T_\odot^{2/3}n_b^{-1/3}e\frac{m_c(m_p+2m_c)}{(m_p+m_c)^{4/3}}\,,
\\
r_p &= T_\odot m_c\,,
 \\
s&=\sin\iota=T_\odot^{-1/3}n_b^{2/3}x_p\frac{(m_p+m_c)^{2/3}}{m_c}\,,
\\
\dot{P}_b&=-\frac{192\pi}{5}T_\odot^{5/3}n_b^{5/3}f(e)\frac{m_p m_c}{(m_p+m_c)^{1/3}}\,.
\end{align}
Here we have defined $T_\odot = GM_\odot/c^3$, $n_b = 2\pi/P_b$ and $f(e)$ is given by
$f(e) = {[1+(73/24)e^2+(37/96)e^4]}/{(1-e^2)^{7/2}}$.

For generic conservative theories of gravity, the post-Keplerian parameters can be written in terms of {\it system independent parameters} as \citep{DamourTaylor:1992}
\begin{eqnarray}\label{eq:cons-ppk}
\dot{\omega}&=&n_b\left(\epsilon-\frac{1}{2}\xi+\frac{1}{2}\right)\frac{\beta^2}{1-e^2}\,,\\
\gamma &=& \frac{e}{n_b}X_B(\GOB+\kappa+X_B)\beta^2\,,\\
r_p&=&\frac{1}{4n_b}X_B\GOB(\epsilon_{0B}+1)\beta^3\,,\\
s_p&=&\frac{n_b x_A}{\beta X_B}\,,
\end{eqnarray}
where we have introduced $\GOB$ which is the ratio between an ``effective''
gravitational constant and the Newton's constant $G$, $\epsilon = 3$ in general
relativity, $\xi = 1$ in general relativity, $\kappa = 0$ in general
relativity and finally $X_A = m_a/M_{tot}$, $X_B = m_b/ M_{tot} =
1-X_A$ and $\beta =(\mathcal{G}M_{tot} n_b)^{1/3}/c$ is a
characteristic velocity for the relative orbital motion. 

In summary, the three models that we consider are: (i) the {\it DD model}~\citep{DamourDeruelle:1986}, in which the Keplerian and post-Keplerian orbital parameters are all considered independent parameters, therefore $\{\vec{\lambda}_\mathrm{K},  \vec{\lambda}_\mathrm{pK} = \{\dot{\omega}, \gamma_p, r_p, s, \dot{P}_b\}\}$; (ii) the {\it DDGR model} in which we assume general relativity is the correct theory of gravity and therefore the parameters are $\{\vec{\lambda}_\mathrm{K}, m_p, m_c\}$; and (iii)
the {\it  DDCG model}~\citep{DamourTaylor:1992}, in which the Keplerian orbital parameters are considered free while the post-Keplerian parameters are constrained to obey the relations imposed by a general metric theory, and therefore $\{\vec{\lambda}_\mathrm{K}, \GOB, \epsilon, \xi, \kappa, \dot{P}_b\}$. 

Given a set of observations from $n$ pulsars $\bd = \{\vec{d}_1,\ldots,\vec{d}_n\}$, our purpose is to infer the parameters $\bl$ according to the predictions of some model $\model$ for the dynamics of the system (and the background information $\binfo$ at hand) \emph{and} to evaluate the probability of the data under the specific model assumption, \emph{i.e.} the \emph{model evidence} or marginal likelihood. 
The posterior density function (PDF) for $\bl$ is obtained via Bayes' theorem:
\begin{align}\label{eq:pdf}
p(\bl|\bd,\model,\binfo)=p(\bl|\model,\binfo)\frac{p(\bd|\bl,\model,\binfo)}{p(\bd|\model,\binfo)}.
\end{align}
Here $p(\bl|\model,\binfo)$ is the prior probability density, $p(\bd|\bl,\model,\binfo)$ is the likelihood function and 
\begin{align}\label{eq:z}
p(\bd|\model,\binfo) = \int\de\bl\, p(\bl|\model,\binfo)p(\bd|\bl,\model,\binfo)
\end{align}
is the \emph{model evidence}.
As we can assume that the observations of different pulsars are statistically independent, the likelihood factorises into the product of each individual likelihood
\begin{align}
p(\bd|\bl,\model,\binfo) = \prod_{i=1}^n p(\vec{d}_i|\bl,\model,\binfo)\,.
\label{eq:like_i}
\end{align}
For observations of a single pulsar, say ``1'', 
the likelihood is simply $p(\vec{d}_1 | \bl,\model,\binfo)$. Given several competing models $\model_{j=1,\ldots,N}$, we can rank their effectiveness in explaining the observed data with the odds ratio:
\begin{align}\label{eq:odds_ratio}
\mathcal{O}_{i,j}\equiv\frac{p(\model_i|\binfo)}{p(\model_j|\binfo)}\frac{p(\bd|\model_i,\binfo)}{p(\bd|\model_j,\binfo)}\,,
\end{align}
where $p(\model_i|\binfo)/p(\model_j|\binfo)$ is the prior odds and 
$p(\bd|\model_i,\binfo)/p(\bd|\model_j,\binfo)$
is the Bayes' factor. Within a given theory (a model or hypothesis) $\model$, Eq.~(\ref{eq:pdf}) provides all the information about the parameters of the theory. If one wishes to compare a given theory with alternatives, the odds ratio Eq.~(\ref{eq:odds_ratio}) is the statistically rigorous quantity to consider \citep{Jaynes:2003}.

%\subsection{Orbital dynamics models}
%\label{ssec:models}

%\subsection{Numerical method}
%\label{ss:numerics}

The likelihood function~(\ref{eq:like_i}) is constructed from the ``timing residuals" $\delta\vec{\tau}$, the difference between the predicted (for a given choice of the parameters of the model) TOAs and the observed TOAs. We take the timing residuals to be distributed according to a multi-variate Gaussian, hence
\begin{equation}\label{eq:like}
p(\bd_i|\vec{\lambda},\model,\binfo) = (2\pi)^{-\frac{k_i}{2}}||\mathbf{C}||^{-1}\,exp{\left[-\frac{^t(\delta\vec{\tau}) \mathbf{C}^{-1} (\vec{\delta\tau})}{2}\right] }\,,
\end{equation}
where $t$ indicates the transpose operation. The expected statistical properties of the residuals are described by the covariance matrix $\mathbf{C}$. $\mathbf{C}$ contains unknown parameters of the model, white and red-noise parameters which too need to be simultaneously estimated. The general form of our likelihood allows for a natural inclusion in the analysis of red noise as a Gaussian Process, {\it e.g.} \cite{vanHaasterenVallisneri:2014}, and of white noise processes as additional diagonal covariance matrices. For the results on simulated data presented in this paper, we ignore these complications and assume that $\mathbf{C}$ is diagonal with elements given by the known measurement uncertainties. Nevertheless, the inclusion of additional red or white noise parameters does not affect our results. 

The marginalised posterior PDFs and the model's evidence are obtained by numerical integrations over a large multi-dimensional parameter space, for which stochastic sampling techniques are particularly suitable. \textsc{TEMPONEST} \citep{LentatiEtAl:2014} provides a specific sampling scheme that can be applied to this problem. We have developed an independent nested sampling algorithm tailored to the problem at hand, that is based on the scheme proposed by \cite{VeitchVecchio:2010}, which is part of the analysis infrastructure to study compact binary systems with gravitational-wave laser interferometers~\citep{LALInference}, and was used to estimate the parameters of the binary black hole merger associated to the transient GW150914 observed by Advanced LIGO and to perform tests of general relativity~\citep{GW150914-DETECTION, GW150914-PARAMESTIM, GW150914-TESTOFGR}. The algorithm (written in \texttt{python}) provides a reliable estimate of the evidence integral as well as the marginalised PDFs on all model parameters. It implements a parallel scheme to generate new samples from the prior probability distribution. The timing model and thus the likelihood are computed through the timing software \textsc{TEMPO2} (\cite{tempo2:1,tempo2:2}) and the python wrapper \texttt{libstempo} (\cite{libstempo}. Any timing model supported by \textsc{TEMPO2} can therefore be immediately used in our code. The results presented in this paper are 
obtained using 1024 ``live points" in the nested sampling -- see for more technical details \cite{VeitchVecchio:2010} -- using standard 8 CPU cores in parallel. Each analysis (details are provided below) takes about 12 hours to complete. This is therefore an efficient and generic approach that can be applied to a large range of observations.

Eq.~(\ref{eq:like}) is, in general, a sharp function of the parameters $\vec{\lambda}$, and % For this reason, 
a search over the whole %entirety of the 
prior volume is inefficient. We thus proceed as suggested in \cite{LentatiEtAl:2014}: from a first analysis of the data with \textsc{TEMPO2} we restrict the domain of integration for the evidence and PDFs calculation to a region of the parameter space which is centred around the \textsc{TEMPO2}'s best fit parameters with a width $\pm 5 \sigma$, where $\sigma$ is the nominal statistical error returned by \textsc{TEMPO2}. As long as the prior volume is large enough to capture the full likelihood function, this strategy is both numerically efficient and formally correct, since the likelihood outside this range does not contribute to the integral. %\footnote{During our testing, we have found that the \textsc{TEMPO2}'s best fit parameters can be several $\sigma$s away from the actual value of $\vec{\lambda}$ (which is always correctly recovered by our Bayesian analysis), and the choice of $\pm 5 \sigma$ is appropriate for the specific cases discussed here. For other observations and systems it needs to be determined as part of the analysis.} 
Finally, we adopt uniform priors for all the parameters in the model.

\section{Results}
\label{sec:Results}

In order illustrate the power of our proposed approach, we present the results from analyses of simulated data of four hypothetical binary pulsar systems. We note that our method is not restricted to Double-Pulsar-like systems, but can operate on any class of binary systems (e.g. neutron star -- white dwarf), thus probing different regions of the space of theories of gravity.  In generating the parameters of the systems under consideration, we assume that general relativity is the correct description of gravity. The double pulsar J0737-3039~\citep{BurgayEtAL:2003} is one of the very best systems to provide tight constraints on theories of gravity, we thus consider a system and data set -- System \#1 in Table~\ref{tab:parameters} -- that qualitatively reproduce the observations reported in~\cite{KramerEtAl:2006}: a three year data set with a cadence of one per month and instrumental timing uncertainty of 20 $\mu$s. The three additional systems and data sets have equal length and cadence as System \#1 and the same (but statistically independent) timing noise. The parameters of the systems are given in Table~\ref{tab:parameters}. The TOAs are generated using the \texttt{fake} package of \textsc{TEMPO2} setting the noise realisation to zero.

%
%%%%%%%%%%%%%%%%%%%%%
\begin{table*}
\caption{The parameters of the simulated binary systems. Additional parameters not shown are the (arbitrary) values of the binary location in the sky -- right ascension $\alpha$ and declination $\delta$ -- the parallax and the TOAs standard deviation $\sigma_\mathrm{TOA}$.}
\centering
\begin{tabular}{l c c c c}
\hline\hline
Parameter		&		System \#1	&	System \#2	&			System \#3	&			System \#4	\\			
\hline
Pulsar 1 spin frequency, $\nu_1$ $(\mathrm{s}^{-1})$								&44.054069292744	&61.678521661049&34.324170685652&0.725902035568\\
Pulsar 1 spin down rate, $\dot{\nu}_1$ $(10^{-14}\mathrm{s}^{-2})$						&-0.34156	&-4.92360&-8.33250&-4.80914\\
Pulsar 2 spin frequency, $\nu_2$  $(\mathrm{s}^{-1})$								&0.36056035506 	&27.532606924712&25.275689037175&8.247581944353\\
Pulsar 2 spin down rate, $\dot{\nu}_2$ $(10^{-14}\mathrm{s}^{-2})$						&-0.01160	&-6.59530&-4.25058&-3.83387\\
Proper motion in $\alpha$, $\mu_\alpha$	$(\mathrm{mas}\,\mathrm{yr}^{-1})$		&-3.3	&-41.5&-5.5&-3.9\\
Proper motion in $\delta$, $\mu_\delta$	$(\mathrm{mas}\,\mathrm{yr}^{-1})$	&2.6	&-4.9&9.2&4.6\\
Dispersion measure, 	$(\mathrm{pc}\,\mathrm{cm}^{-3})$							&48.920	&11.1327&8.471&1.595\\
Orbital period, $P_b$ $(\mathrm{days})$ 														& 0.10225156248 	&0.70853013025&0.09536840065&0.04495305403\\	
Projected semi-major axis 1, $a_1$	$(\mathrm{lt-s})$											& 1.41094774993 	&3.37904426006&1.51878410152&0.90797440542\\
Projected semi-major axis 2, $a_2$	$(\mathrm{lt-s})$											& 1.51172166241 	&4.32700630238&1.20719408862&0.80919560400\\
Time of periastron passage, $T_0$ (MJD)												& 53155.9074280	&53252.2188315&53436.2200040&53892.2676109\\
Angle of periastron, $\omega_1$ (deg)														& 87.0331	&155.3942&77.0200&20.6494\\
Orbital eccentricity, $e$																		& 0.0877775	&0.8754259&0.1538017&0.0120680\\
Mass of pulsar 1, $m_1$ $(M_\odot)$														& 1.3381	&1.0504&1.1533&1.2986\\
Mass of pulsar 2, $m_2$ $(M_\odot)$														& 1.2489	&1.3451&1.4510&1.4571\\
\hline
\end{tabular}
\label{tab:parameters}
\end{table*}
\begin{figure*}
\includegraphics[width=2\columnwidth]{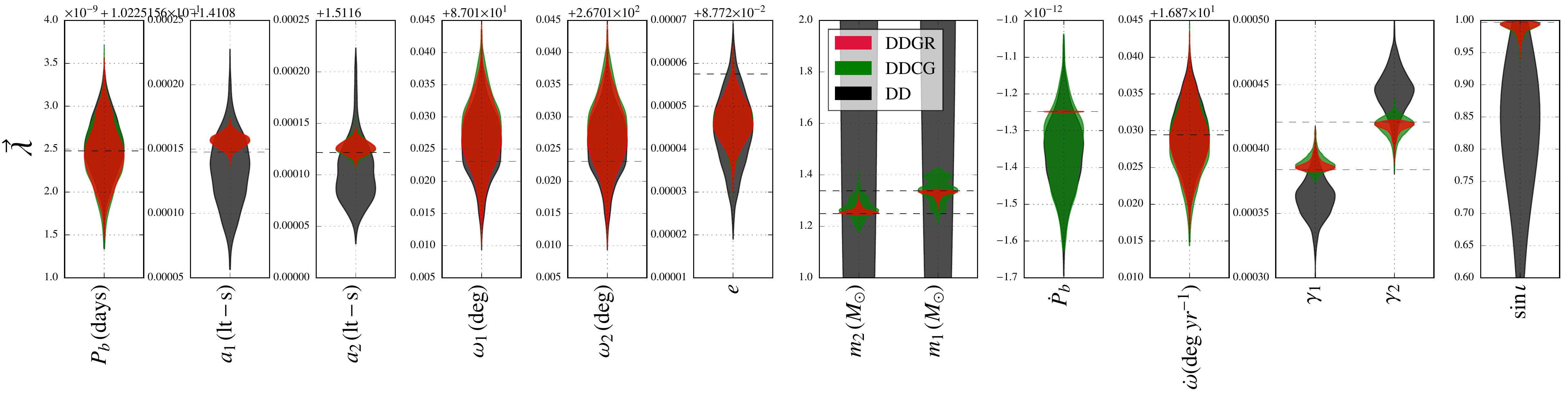}
\caption{Violin plot summarising the posterior distributions for the Keplerian and post-Keplerian parameters from the analysis of System \#1 using the DD (black), DDGR (red) and DDCG (green) models. In each panel the thin dashed line indicates the value of the corresponding parameter used to generate the simulated data. We note the non-Gaussian nature of the PDFs for the mass parameters as well as for $\gamma_1$ and $\gamma_2$, highlighting the failure of the Laplace quadratic approximation of the likelihood function.}
\label{fig:violin-pos}
\end{figure*}

\begin{figure*}
\includegraphics[width=0.9\columnwidth]{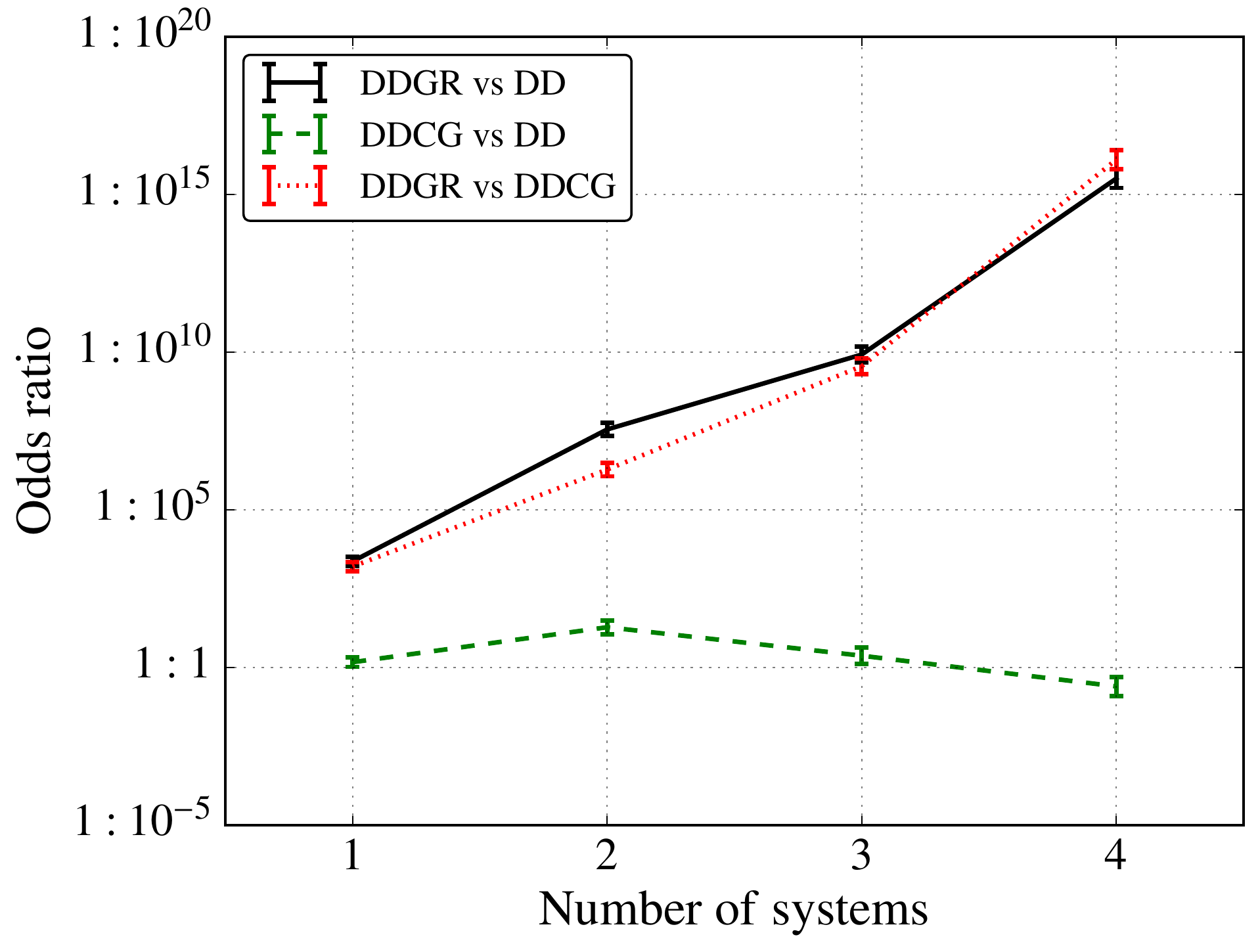}
\includegraphics[width=\columnwidth]{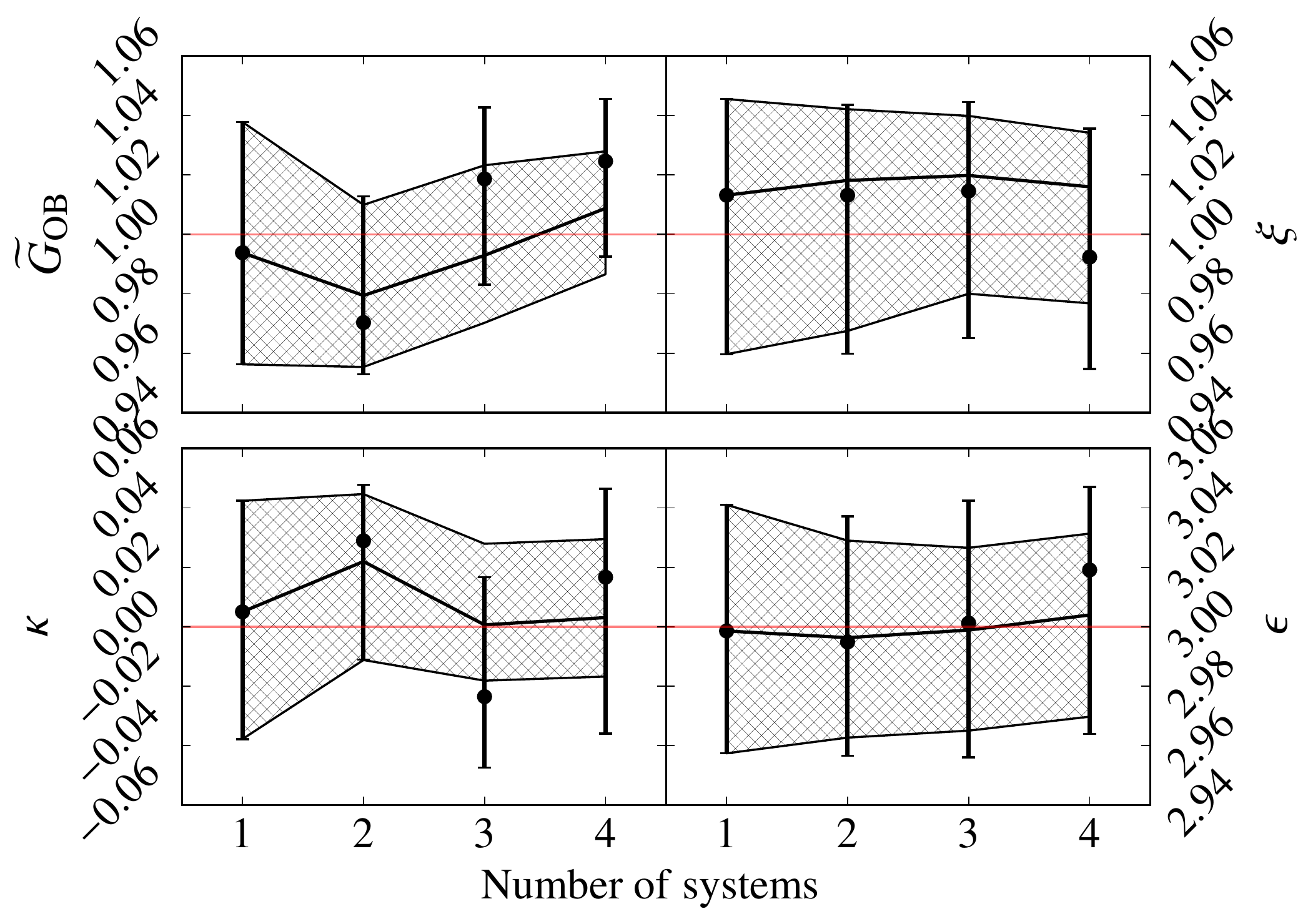}
\caption{
%Summary of the main results in this paper. {\it Top}: Examples of two-dimensional marginalised PDFs on selected model parameters for system 1. -- left: component masses $(m_A, m_B)$; right: post-Keplerian parameters $(\dot{\omega}, \dot{P}_b)$ -- using the DD (black), DDGR (red) and DDCG (green) models. Points correspond to posterior samples and lines to the 67\%, 95\% and 99\% central probability intervals obtained with a Dirichlet Process density estimator \citep{bnp}. {\it Bottom}: 
Results from the joint analysis of the four systems. Left panel: cumulative odds ratios between the different models as a function of the number of systems included in the analysis. The error-bars indicate the theoretical uncertainty in the odds due to the discrete approximation of Eq.~(\ref{eq:z}). The average uncertainty is 0.2 dex. Right panel: 95\% probability interval computed from the joint PDF (hatched regions) for the global parameters $\GOB$ (top left), $\epsilon$ (top right), $\xi$ (bottom left) and $\kappa$ (bottom right) from the analysis on one (dashed), two (dotted), three (dot-dashed) and four (solid) systems. The  solid horizontal line indicates the value predicted by general relativity, which has been used to generate the simulated observation data.}
\label{fig:CG-pos}
\end{figure*}

We illustrate our findings by discussing three key points. Firstly, we show how constraints on general relativity can be rigorously quantified in terms of odds ratios between different models. Secondly, we show that the joint inference from the observations of many systems can yield more stringent evidence for --or against -- a specific theory.
Finally, constraints on system independent parameters that are characteristic of the theory improve when one combines the data from different systems. The main result of our analysis are summarised in Figs.~\ref{fig:violin-pos} and~\ref{fig:CG-pos}, and are discussed in detail below.

We analyse the data using the three models, DD, DDGR, and DDCG discussed in 
Section \ref{sec:ModelMethod}. The number of parameters that describe the models is 24, 18 and 24, respectively. All the models are described by the {\it same set} of non-orbital parameters $\vec{\theta} = \{\nu_A, \dot{\nu}_A, \nu_B, \dot{\nu}_B, \alpha, \delta, \mu_a, \mu_\delta, D\}$ and Keplerian parameters $\vec{\lambda}_\mathrm{K} = \{P_b, x_a, x_b, e, \omega, T_0\}$. In addition for the DDGR model we have the parameters $\lambda_\mathrm{pK} = \{m_A, m_B\}$,
for the DD model the parameters $\lambda_\mathrm{pK} = \{\dot{\omega}, \dot{P}_b, \gamma_A, \gamma_B, r_A, r_B\}$ and for the DDCG model the parameters  $\lambda_\mathrm{pK} = \{\GOB,\epsilon,\kappa,\xi\}$. We then proceeded with the calculation of the PDFs for all parameters as well as the evidence for the three models under consideration. With four independent systems the total number of unknown parameters in the analysis is 96, 72 and 96 for the DD, DDGR, and DDCG model, respectively.  Fig.~\ref{fig:violin-pos} shows one-dimensional marginalised PDFs for System \#1 on
$\lambda_\mathrm{K}$, $\lambda_\mathrm{pK}$, $m_1$ and $m_2$ under the three different model assumptions. The simulated values of the parameters are correctly recovered.  Similar PDFs (not shown) are obtained for the remaining Systems. The widths and shapes of the PDFs vary considerably depending on the assumed orbital model. Our approach and its numerical implementation provides therefore a powerful tool to derive rigorous statistical information from pulsar timing.

Fig.~\ref{fig:CG-pos} provides an example of the comparison of the models and the inference of global parameters of the theory when jointly analysing the data from multiple systems. The data were generated according to the DDGR model, and analysed in turn under the three models assumptions, DD, DDGR, and DDCG. Indeed, the results of the analysis shows that the DDGR model is vastly favoured, and as more systems are included in the analysis this conclusion becomes progressively stronger. The left panel of Fig.~\ref{fig:CG-pos} shows this result in a striking way. From a single system, the odds in favour of the DDGR model are ${\cal O}_\mathrm{DDGR, DD}  = 2.3\times 10^3 : 1$ and ${\cal O}_\mathrm{DDGR, DDCG} = 1.6\times 10^3 : 1$ against the DD and DDCG model, respectively; the DDGR model is $\sim 1000$ times more probable than either alternatives. With the inclusion of more systems in the analysis the odds in favour of DDGR increase substantially, reaching values of $3\times 10^{15} : 1$ and $10^{16} : 1$, respectively. These results provide also a qualitative indication of how much more stringent the results from J0737-3039 could become as the time-span of the observations increases (assuming that the physics at work does follows the DDGR model).  As for the DDCG model, the odds do not show any significant evidence in favour of DDCG against DD. From the analysis of one system DDCG shows odds against DD that are $1:1.5$ which from the analysis of 4 systems become $1:1/4$. These values are consistent with the numerical errors of the evidence evaluation. Correctly, the data do not provide evidence in favour of DDCG against DD.

The analysis of the data using the DDCG model, described by system independent parameters whose value in GR and other theories of gravity is known \emph{a priori}~\citep[e.g.][]{KramerWex:2009} --  $\GOB$, $\epsilon$, $\xi$ and $\kappa$ -- allows us to construct joint PDFs on these parameters, which in turns shows another advantage of our method. More strict limits on system-independent parameters can be obtained by a joint analysis of several systems. We stress that we perform the analysis by \emph{simultaneously} fitting for all the parameters in the model for all the sources, and not by fixing the values of some (or all the other parameters) to a some arbitrary value (say the maximum likelihood or best fit value) and then estimating these global parameters. The right panel of Fig.~\ref{fig:CG-pos} shows the median of each of the parameters and the central 95\% probability interval coming from the {\it independent} analysis of each system, and the contraction of the 95\% probability interval produced in the {\it joint} analysis.
The probability intervals both for the individual system analyses and the joint analysis correctly contain the GR values, which were used to generate the data sets. We find that as systems are added to the analysis the size of the probability interval decreases roughly as the square root of the number of systems considered. Furthermore, for the specific parameters of the simulated data that we consider here, one can reach an accuracy of a few percent at the $95\%$ confidence level on all the parameters. We note that, in reality, the measurement accuracy will also depend on the length of the observations and scale differently depending on the post-Keplerian parameter.

\section{Conclusions}
\label{sec:Conclusions}

We have introduced a Bayesian approach to perform tests of general relativity using timing data of binary radio pulsars.  It relies on the evaluation of the odds ratio between different theories on a system-by-system basis and in addition provides more stringent constrains by \emph{combining} the results from the observation of multiple systems. If the specific theory of gravity that one wishes to test is described by global parameters (\emph{i.e.} independent of the specific binary system), this approach also provides constraints on these parameters coming from the joint analysis of all the systems at hand. We have illustrated its the power via an efficient numerical implementation (which is appropriate for the large number of dimensions that characterise this problem) applied to a synthetic data set representative of plausible observations of several binary pulsars. This approach can and should be applied to existing and future data sets.

\textit{Acknowledgments---}
We thank Michael Kramer, Michele Vallisneri and John Veitch for useful comments and discussions. The work was funded
in part by a Leverhulme Trust research project grant. The numerical simulations were performed
on the Tsunami cluster of the University of Birmingham.
\bibliographystyle{mnras}
\bibliography{bibliography} 

\label{lastpage}

\end{document}